# A toolbox for general elliptical gravitational lenses

T. Schramm

Hamburger Sternwarte, Gojenbergsweg 112, D-2050 Hamburg 80, Germany

**Abstract.** We introduce a formalism for constructing models of the gravitational lens action of arbitrary elliptical surface mass distributions including varying ellipticity and isodensity-twist. The *projected* elliptical surface mass distributions are cut to slices of constant surface mass density for which the deflection is easy to compute. The deflection of the total mass distribution is found by superposition of the action of all slices.

## 1. Introduction

There is an ongoing discussion on how to model elliptical mass distributions like elliptical galaxies and clusters as gravitational lenses. For these models the 2-dimensional potential of the *projected* mass distribution is needed. For some elliptical 3-dimensional mass distributions (as, e. g., triaxial, homoeoidal) the isodensities of the projected surface mass distribution are exact elliptical curves. For others (e. g. confocal) this is only approximately true (see e. g. Schneider et al. 1992 or Schramm 1990, hereafter Paper I).

Some authors prefer to start with 2-dimensional elliptical potentials which are easy to compute (Blandford & Kochanek 1987, Kochanek & Blandford 1987). Their disadvantage is that the related surface mass distributions do not share their symmetries and that they are not applicable for mass distributions with high ellipticity.

Other authors prefer to start with elliptical mass distributions. Bourassa et al. (1975, hereafter BK) introduced a complex formalism describing the lens action of projected spheroidal (and homoeoidal) mass distributions with arbitrary ellipticity. Schramm (Paper I) found that their results are also valid for projected triaxial distributions and introduced an equivalent real formalism. Closed solutions have rarely been found and are by construction restricted to the *homoeoidal* symmetry, where the isodensities are similar and similarly situated, concentric ellipses (see Kassiola & Kovner 1993, hereafter KK, for recent results or Narasimha et al. 1982 and below).

Both approaches differ only slightly concerning image configurations produced by elliptical lenses of moderate ellipticity (see also KK). However, using potentials of elliptical surface mass distributions allows more freedom in modelling elliptical galaxies and clusters. On the one hand some special elliptical surface mass distributions become tractable for theoretical purposes, on the other hand the isophotes of observations of elliptical galaxies are typically fitted by ellipses using standard packages like STSDAS. The results are three functions of the *elliptical radius* (introduced below), namely intensity, ellipticity and isophote-twist which may be used as input for modelling gravitational lenses (see e.g. Bernstein et al. 1993). Even if the mass-to-light ratio is not known, the mass might share the symmetry of the isophotes (compare e.g. Peletier 1989) and therefore only the mass-to-light ratio as a function of radius must be fitted. A failure of this fit would indicate different symmetries for light and mass.

Our program is to take the 2-dimensional, projected surface density distribution with elliptical symmetry and cut it into slices of constant density (see Fig. 1). This means that we substitute the density distribution by a step function. For approximative purposes this function could be chosen as the lower, upper or symmetric case of the step function where the position of the steps are chosen as given by observations or computational needs. (In Fig. 1 we show the lower step function as an example.) For analytical purposes the choice of step function does not matter as long the density is Riemann integrable. We therefore define an *elliptical slice* to be a bounded elliptical area of constant surface mass density. (We avoid the expression *elliptical disk* which could be misleading.) For these elliptical slices we present solutions for the lens action including potentials (for the time delay) and derivatives (to find the critical and caustic structure of the lens mapping). It is then shown how to superpose elliptical slices to obtain a description for arbitrary elliptical surface density distributions. If the slices are taken to be of finite thickness, we obtain an approximate, numerical de-scription. An appropriate, exact analytical description is found by taking the limit of infinitely thin slices and integrating over all their contributions.

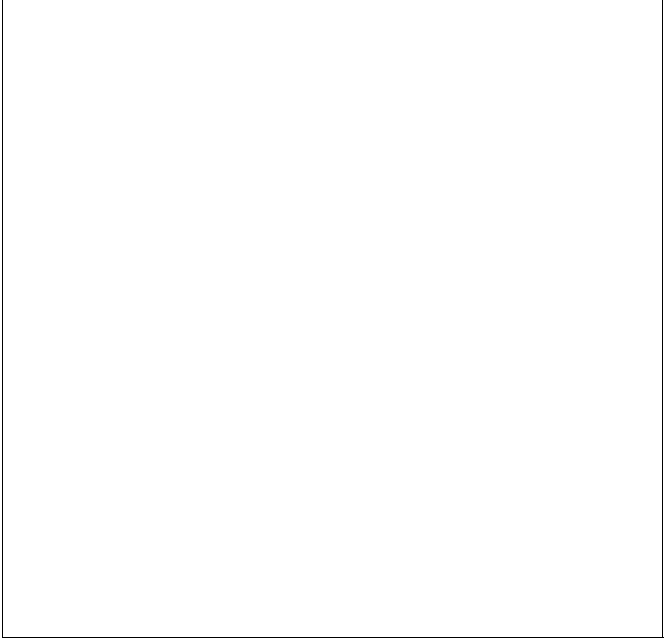

**Fig. 1.** The surface mass density $\sigma(r_E)$ as a function of the elliptical radius defined in Eq.(6) in arbitrary units. Also given is a related lower step function $s(r_E)$. The shaded region represents an elliptical slice of thickness $\triangle\sigma$

## 2. Basics

The gravitational lens action of a deflector at angular size distance $D_d$ on a source at distance $D_{ds}$ from the deflector and $D_s$ from the observer is described by

$$w(z, \bar{z}) = z - \nabla \Phi(z, \bar{z}), \tag{1}$$

where the complex quantities $w = u + iv, z = x + iy$ denote angular positions in the source and deflector plane, respectively. The bar denotes complex conjugation and the gradient is defined by $\nabla = \partial/\partial x + i\partial/\partial y = 2\partial/\partial \bar{z}$.

Although we introduced in Paper I a real formalism for the purpose of making clear the geometrical meaning of the occurring quantities, we use here the complex notation for several reasons: (a) it is typically not necessary to split the lens equation into real and imaginary parts unless we need numerical data, which is normally automatically achieved using complex variables in e.g. FORTRAN; (b) some results are found in a very simple and elegant form, e.g. the lens action of an elliptical slice or the parametric equations for the critical curves as introduced by Witt (1990); (c) if it is unavoidable to look at real and imaginary parts for analytical purposes, the equations can be transformed using powerful symbolical algebra systems like MAPLE$^{TM}$. We used an implementation of this software on an AMIGA 3000$^{TM}$ computer system for evaluating and checking our formulas and for preprocessing the figures.

The real potential $\Phi$ obeys the Poisson equation

$$\triangle \Phi(z, \bar{z}) = 2\sigma(z, \bar{z}), \tag{2}$$

where the dimensionless real surface density is defined as

$$\sigma(z, \bar{z}) = \frac{\Sigma(z, \bar{z})}{\Sigma_c} \tag{3}$$

and the critical surface mass density is as usual defined by

$$\Sigma_c = \frac{c^2}{4\pi G} \frac{D_s}{D_d D_{ds}} \quad, \tag{4}$$

where all distances $D_{ij}$ are measured as apparent size distances.

An ellipse is defined by

$$\frac{x^2}{a^2} + \frac{y^2}{b^2} = 1 \quad, \text{or} \tag{5}$$

$$\frac{x^2}{(1+\epsilon)^2} + \frac{y^2}{(1-\epsilon)^2} = r_E^2 \tag{6}$$

respectively. The parameters $a, b$ are the major and minor axis of the ellipse, which is evidently also fully determined by the elliptical parameter $\epsilon$ and the *elliptical radius* $r_E$ (note that the $\epsilon$-parameter is neither equal to the ellipticity which is $1 - b/a$, nor equal to the numerical eccentricity which is $\sqrt{a^2 - b^2}/a$, but is related to both). For convenience we give some useful relations between both descriptions Eqs. (5) and (6):

$$\epsilon = \frac{a-b}{a+b} \quad, \tag{7}$$

$$r_E = \frac{a+b}{2} \quad, \tag{8}$$

and for the foci $\pm f$ of the ellipse

$$f^2 = a^2 - b^2 = 4\epsilon r_E^2. \tag{9}$$

Note that the relation between $f$ and $r$ is restricted by $0 \le \epsilon \le 1$ so that for a constant $f$ a lower limit $r_{min} = f/2$ exists.

A surface mass distribution $\sigma = \sigma(r_E)$ with $\epsilon = $ const. is called *homoeoidal*, and with $f = $ const. it is called *confocal*. The isodensities due to a confocal distribution are given by a set of confocal ellipses, parameterized by $\lambda$, confocal with respect to the ellipse given by Eq.(5)

$$\frac{x^2}{a'^2} + \frac{y^2}{b'^2} = \frac{x^2}{a^2 + \lambda} + \frac{y^2}{b^2 + \lambda} = 1 \quad. \tag{10}$$

For given $a, b$ we find the confocal ellipse at $x, y$ solving Eq. (10) for $\lambda$.

Elliptical slices acting as gravitational lenses are interesting for several reasons. As we shall see they represent inside a *pure* $\sigma - \gamma$ (convergence-shear) lens with two critical densities (for a constant density: two perpendicular focal lines at different distances). Taking one slice alone (e.g. as a cluster model), it could therefore produce very elongated images of a source. A detailed discussion of this lens type will be published in a subsequent paper (Schramm & Kayser 1993). For our purposes elliptical slices will be the basic element of our construction set.

We derive the lensing properties of an elliptical slice as a trivial consequence of the results obtained in BK. They could also be found following the track of the Appendix A in Paper I (see below). Blandford et al. (1991) obtained later an equivalent result for the lens action of an inclined disk using elliptical coordinates (compare also Appendix B of Paper I).

### 3.1. A short summary of BK's results

BK introduced a formula describing the lens action of a homoeoidal mass distribution

$$\bar{I}(z, \bar{z}) = 2(1 - \epsilon^2)\text{sign}(\bar{z}) \int_0^{r_{\text{E}}(z)} \frac{\sigma(r) r \, dr}{\sqrt{\bar{z}^2 - 4\epsilon r^2}}, \qquad (11)$$

where $r_{\text{E}}(z)$ is given by Eq. (6). We note that we integrate over all matter inside the ellipse hit by $z$. The outer parts do not contribute to the deflection. The integrand could be interpreted as the lens action due to an infinitesimal thin homoeoidal ring of constant surface mass density. For details see the Appendix B of Paper I. A generalized signum function sign($z$) was introduced by Bray (1984) to assure the right symmetry behaviour of the BK formula. The definitions given below assure the equivalence to the results obtained with real formalism of Paper I, if the canonical choice for the principal value (as taken by FORTRAN77 compilers) is chosen for the evaluation of complex square roots. Other choices are possible, however. Bray's definition also ensures that the internal and external solutions obtained below (Eq. (15), Eq. (17)) coincide at the boundary of the slice.

Note that the following definition differs from the canonical signum function as typically implemented in computer (algebra) languages. Bray defined:

$$\begin{aligned}\text{sign}(z) &= \frac{\sqrt{z^2}}{z} \qquad (12) \\ &= \begin{cases} +1 & \text{if } x > 0 \text{ or } (x = 0 \text{ and } y \geqq 0) \\ -1 & \text{otherwise} \end{cases}. \end{aligned}$$

Note that sign($\bar{z}$) changes only the sign of the $y$-axis, so that

$$\text{sign}(z)\text{sign}(\bar{z}) = \begin{cases} -1 & \text{if } x = 0 \\ +1 & \text{otherwise} \end{cases} \qquad (13)$$

$$\nabla \Phi = \bar{I} \qquad (14)$$

### 3.2. The deflection due to an elliptical slice

For $\sigma = \sigma_0 = $ const. Eq.(11) is easily evaluated. We describe the lens action of an elliptical slice with elliptical radius $r_{\text{E}}$. There is an *external* solution for $z$ lying outside

$$\begin{aligned} \bar{I}_{\text{ex}} &= -\frac{1 - \epsilon^2}{2\epsilon}\text{sign}(\bar{z})\left(\sqrt{\bar{z}^2 - 4\epsilon r_{\text{E}}^2} - \sqrt{\bar{z}^2}\right)\sigma_0 \\ &= \frac{1 - \epsilon^2}{2\epsilon}\left(\bar{z} - \text{sign}(\bar{z})\sqrt{\bar{z}^2 - 4\epsilon r_{\text{E}}^2}\right)\sigma_0 \qquad (15) \\ &= 2\frac{ab}{f^2}\left(\bar{z} - \text{sign}(\bar{z})\sqrt{\bar{z}^2 - f^2}\right)\sigma_0 \qquad (16) \end{aligned}$$

and an *internal* solution for $z$ lying inside. We note that we have only to integrate up to the elliptical radius corresponding to $z$. With $\bar{z} = x - \mathrm{i}y$ and Eq. (6) it follows that (compare Eq. (A3) of KK)

$$\begin{aligned} \bar{I}_{\text{in}} &= ((1 - \epsilon)x + \mathrm{i}(1 + \epsilon)y)\sigma_0 \\ &= (z - \epsilon\bar{z})\sigma_0 \quad . \qquad (17)\end{aligned}$$

Note that $\bar{I}_{\text{in}}$ does not depend on the actual size of the slice but on the elliptical parameter $\epsilon$ defined by the bounding ellipse. Equations (15-17) are equivalent to the results obtained in Appendix A in Paper I. The deflection angle $\alpha$ due to an elliptical slice reads with our normalization:

$$\alpha = \alpha_x + \mathrm{i}\alpha_y = -2\frac{ab}{a' + b'}\left(\frac{x}{a'} + \mathrm{i}\frac{y}{b'}\right)\sigma_0 \quad , \qquad (18)$$

where the primes denote the confocal semi-axes defined by Eq.(10). For a proof of the equivalence one has to show that

$$\alpha = -\bar{I} \quad . \qquad (19)$$

For the internal solution this is a trivial task since in Eq. (18) the primes are dropped. For the external solution one can show after some algebra that Eq. (19) leads to Eq. (10). We repeat the result from Paper I since in the real version it is easier to see that *the norm of the deflection has homoeoidal symmetry inside and confocal symmetry outside the deflecting slice*. In addition it is easy to see that with increasing $|z|$ the deflection angle ( in this confocal region) converges monotonically to the deflection angle of a point mass.

### 3.3. Derivatives and critical structures

Calculus of non-analytic complex functions could be expressed in terms of Wirtinger-derivatives (see e.g. Jänich,

complex function $F$ are

$$dF = \frac{\partial F}{\partial z}dz + \frac{\partial F}{\partial \bar{z}}d\bar{z} \qquad (20)$$

$$\frac{\partial F}{\partial z} = \frac{1}{2}\left(\frac{\partial F}{\partial x} - i\frac{\partial F}{\partial y}\right), \quad \frac{\partial F}{\partial \bar{z}} = \frac{1}{2}\left(\frac{\partial F}{\partial x} + i\frac{\partial F}{\partial y}\right) \qquad (21)$$

$$\frac{\partial F}{\partial x} = \frac{\partial F}{\partial z} + \frac{\partial F}{\partial \bar{z}}, \quad \frac{\partial F}{\partial y} = i\left(\frac{\partial F}{\partial z} - \frac{\partial F}{\partial \bar{z}}\right) \qquad (22)$$

$$\frac{\partial \bar{F}}{\partial z} = \overline{\frac{\partial F}{\partial \bar{z}}}, \quad \frac{\partial \bar{F}}{\partial \bar{z}} = \overline{\frac{\partial F}{\partial z}} \qquad (23)$$

For the elliptical slice these derivatives become very simple since $\partial \bar{I}_{\mathrm{ex}}/\partial z = 0$ outside and $\partial \bar{I}_{\mathrm{in}}/\partial z = \sigma_0$ inside (compare also Witt 1990 and Schramm et al. 1993 for details). The derivatives of the components of the deflection angle with respect to the coordinates $(x, y)$ can be found by using Eq. (22) and splitting it in its real and imaginary parts.

The determinant of the Jacobian of gravitational lens mappings $w$ is given by

$$J = \left(\frac{\partial w}{\partial z}\right)^2 - \frac{\partial w}{\partial \bar{z}}\overline{\frac{\partial w}{\partial \bar{z}}} \quad , \qquad (24)$$

since $\partial w/\partial z$ is always a real function. For the lens mapping due to an elliptical slice we find for $J$ in the inner region simply

$$J_{\mathrm{in}} = (1 - \sigma_0)^2 - (\epsilon\sigma_0)^2 \qquad (25)$$

which shows that we find two focal lines, $J = 0$, as typical for pure *astigmatic* lenses, corresponding to two critical values for the density

$$\sigma_{1,2} = \frac{1}{1 \pm \epsilon} \qquad (26)$$

For the external case we note that $\bar{I}_{\mathrm{ex}}$ in Eq. (15) is an (anti)analytical function solely of $\bar{z}$ (since there is no matter outside the slice). Then the parametric form of the critical curve $J_{\mathrm{ex}} = 0$ (Witt 1990) becomes

$$e^{i\phi} = -\frac{\partial \bar{I}}{\partial \bar{z}} \quad , 0 \leq \phi \leq 2\pi \qquad (27)$$

which leads to a parametric equation of the external critical curves

$$\bar{z} = \frac{f\left(e^{i\phi} + \frac{1-\epsilon^2}{2\epsilon}\sigma_0\right)}{\sqrt{e^{i\phi}(e^{i\phi} + \frac{1-\epsilon^2}{\epsilon}\sigma_0)}} \quad . \qquad (28)$$

Note that we have to take only those parts of these curves which lie inside (for Eq. (25)) or outside (for Eq. (28)) of the considered elliptical slice, respectively.

In matter free regions the varying part of the potential is given by (compare also Appendix B in Paper I)

$$\Phi_{\mathrm{ex}} = \mathrm{Re}\int I_{\mathrm{ex}}(z)dz \quad , \qquad (29)$$

which yields

$$\Phi_{\mathrm{ex}} = \mathrm{Re}\frac{1-\epsilon^2}{4\epsilon}\left(f^2\ln\left(\frac{\mathrm{sign}(z)z + \sqrt{z^2 - f^2}}{2}\right)\right.$$
$$\left. - \mathrm{sign}(z)z\sqrt{z^2 - f^2} + z^2\right)\sigma_0 \qquad (30)$$

The constant of integration was chosen to achieve a variation as $\ln|z|$ far from the slice. Inside the slice the potential is given by

$$\Phi_{\mathrm{in}} = \int \mathrm{Re}\left(\bar{I}_{\mathrm{in}}\right)dx + \int \mathrm{Im}\left(\bar{I}_{\mathrm{in}}\right)dy \qquad (31)$$

which is simply

$$\frac{1}{2}\left((1-\epsilon)x^2 + (1+\epsilon)y^2 +\right)\sigma_0 + C \quad , \qquad (32)$$

where the constant $C$ is chosen as

$$C = \sigma_0 r_{\mathrm{E}}^2(1-\epsilon^2)\ln(r_{\mathrm{E}}) \quad . \qquad (33)$$

With this choice, in order to give a continuous variation across the boundary of the slice, the potential is smooth and the boundary of the slice turns out to be only a point of inflection. As shown in Fig.(2) the potential shows interior a homoeoidal symmetry but with lower eccentricity as the slice. Far from the slice it tends towards the potential of a point mass.

*3.5. Rotated and shifted slices*

We can express the lens action of an elliptical slice which is rotated by an angle $\varphi$ with respect to the $(x, y)$–frame. The relation for the lens action between the original and the rotated frame $(x', y')$ reads:

$$\bar{I} = e^{i\varphi}(\bar{I})' \quad ; \qquad (34)$$

note that in general $(\bar{I})' \neq \overline{(I')}$. With $z' = ze^{-i\varphi}$ and $\overline{(z')} = \bar{z}e^{i\varphi}$ we find in the external region

$$\bar{I}_{\mathrm{ex}} = \frac{1-\epsilon^2}{2\epsilon}\left(\bar{z}e^{2i\varphi} - e^{i\varphi}\mathrm{sign}\left(\bar{z}e^{i\varphi}\right)\sqrt{\bar{z}^2e^{2i\varphi} - f^2}\right)\sigma_0 (35)$$

and in the internal region

$$\bar{I}_{\mathrm{in}} = \left(z - \epsilon\bar{z}e^{2i\varphi}\right)\sigma_0 \quad . \qquad (36)$$

Note that the phase factor $e^{i\varphi}$ should not be absorbed in the square root unless its sign is taken into account.

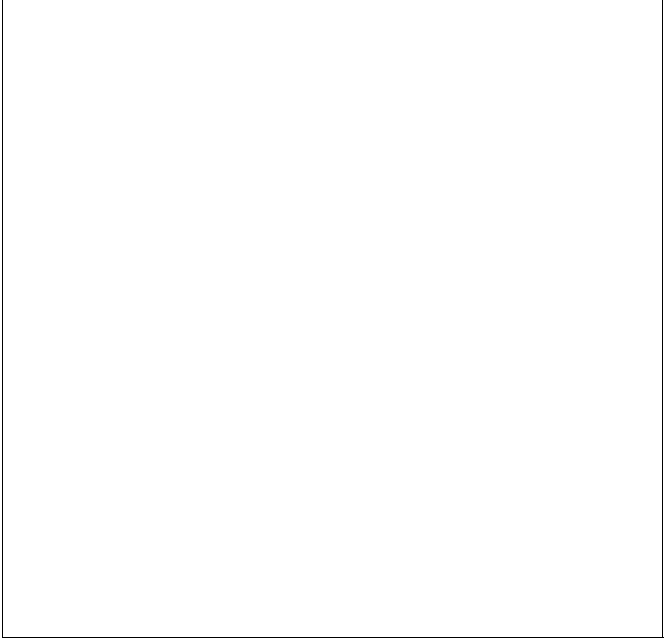 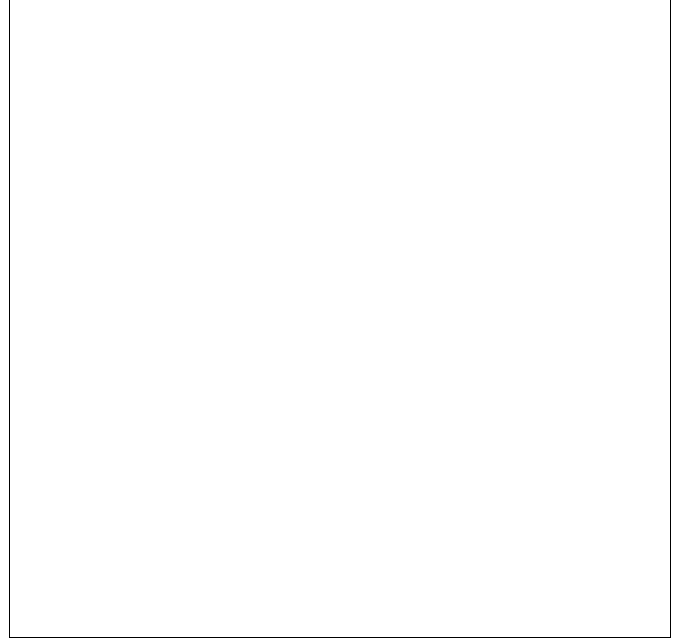

**Fig. 2.** A contour – grey-scale plot of the potential of the slice which is indicated by the bold white ellipse.

**Fig. 3.** The norm of the deflection angle for a rotated ($\varphi = 30°$) slice with major and minor axis $a = 2, b = 1$ is shown. The shading increases with the norm (dark $\simeq$ high value of the norm). The iso-lines show line of constant norm (white) and argument (black) respectively. Note the homoeoidal symmetry inside the slice and the confocal symmetry outside. The argument of the complex deflection is a linear function inside and turns to a hyperbola outside.

Typically we deal with concentric isodensities. However, we can describe also shifted slices by a proper transformation to the new origin $z_0$ of the shifted slice. We find simply

$$\bar{I} = \bar{I}(z - z_0, \bar{z} - \bar{z}_0) \quad . \tag{37}$$

Obviously Eq. (35-36) and Eq. (37) can be combined to describe rotated and shifted slices. In Fig.(3) we show the deflection angle due to a rotated slice of constant density with major and minor axis $a = 2$, $b = 1$ respectively. The shading increases with increasing norm of deflection. The iso-lines show the homoeoidal symmetry (with same axial ratio as the slice) inside and the confocal symmetry outside the slice.

Potentials and derivatives of rotated slices can be found if the related equations are transformed as shown in this section.

### 4. Elliptical models for numerical simulations

The density distributions of gravitational lenses are not directly measurable. However, making special assumptions about the mass-to-light variation or taking other sources of information (velocity dispersion, theoretical models ...) projected, elliptical density distributions are appropriate for modeling elliptical lenses, i. e. the data obtained from observation can be used as direct input to the model. Analyzing elliptical galaxies typically yields a set of $n$ elliptical isophotes (isodensities) $\sigma_i$ with elliptical radius $r_{Ei}$, their elliptical parameter $\epsilon_i$, their foci $f_i$ and the position angle of their major axis (twist angle) $\varphi_i$. We find the deflection $\bar{I}^{tot}$ at a location $z$ by a summation over all slices of thickness $\triangle \sigma_i = \sigma_i - \sigma_{i-1}$.

$$\bar{I}^{tot} = \bar{I}^{tot}_{in} + \bar{I}^{tot}_{ex} \tag{38}$$

All *internal* contributions are found if

$$r(z) = \sqrt{\left(\frac{\mathrm{Re}\,(ze^{-\mathrm{i}\varphi_i})}{1 + \epsilon_i}\right)^2 + \left(\frac{\mathrm{Im}\,(ze^{-\mathrm{i}\varphi_i})}{1 - \epsilon_i}\right)^2} < r_{Ei} \tag{39}$$

from Eq.(36)

$$\bar{I}^{tot}_{in} = \sum_i \bar{I}_{in}(\triangle\sigma_i, \epsilon_i, \varphi_i) \tag{40}$$

and all *external* contributions for these slices for which $r(z) > r_{Ei}$ from Eq.(35)

$$\bar{I}^{tot}_{ex} = \sum_i \bar{I}_{ex}(\triangle\sigma_i, f_i, \varphi_i) \tag{41}$$

Derivatives and potentials can be found respectively.

Let us take an elliptical density distribution described by the set of consistent functions introduced before and possibly bounded by an ellipse with radius $r_{\max}$

$$\sigma = \sigma(r_E) \quad ,$$
$$\epsilon = \epsilon(r_E) \quad ,$$
$$\varphi = \varphi(r_E) \quad ,$$

and in addition

$$f = f(r_E) \quad . \tag{42}$$

In the limit of infinitely thin slices $\triangle \sigma \to d\sigma$ introducing the derivative of $\sigma$, we replace

$$d\sigma(r_E) = \sigma'(r_E) dr_E \tag{43}$$

to find the limit of Eq.(38) using condition Eq.(39) and $r_{\min} = 0$ (compare note after Eq. (9))

$$\bar{I}_{\text{ex}}^{\text{tot}} = -\int_{r_{\min}=0}^{r(z)} dr_E \, \sigma'(r_E) \frac{1-\epsilon(r_E)^2}{2\epsilon(r_E)} \left[ \bar{z} e^{2i\varphi(r_E)} \right.$$
$$\left. - e^{i\varphi(r_E)} \text{sign}\left(\bar{z} e^{i\varphi(r_E)}\right) \sqrt{\bar{z}^2 e^{2i\varphi(r_E)} - f(r_E)^2} \right] \tag{44}$$

$$\bar{I}_{\text{in}}^{\text{tot}} = -\int_{r(z)}^{r_{\max}} dr_E \, \sigma'(r_E) \left( z - \epsilon(r_E) \bar{z} e^{2i\varphi(r_E)} \right) \tag{45}$$

and

$$\bar{I}^{\text{tot}} = \bar{I}_{\text{in}}^{\text{tot}} + \bar{I}_{\text{ex}}^{\text{tot}} + \bar{I}(\sigma(r_{\max})) \quad , \tag{46}$$

where $\bar{I}(\sigma(r_{\max}))$ is the contribution of the slice of thickness $\sigma(r_{\max})$ which is obviously zero if we deal with *unbounded* densities which decrease to zero at infinity. This term could be omitted if the proper derivatives of the density are invoked at its steps but for practical purposes we keep it and take simply the proper one-sided derivatives. For bounded distributions we have to take the internal or external solution (Eq. (17) or Eq. (15)) for that single slice if the location of $z$ is inside or outside of the the distribution, respectively. For the latter case we note $\bar{I}_{\text{in}}^{\text{tot}} = 0$ since $\sigma' = 0$ in that region. The remaining integral $\bar{I}_{\text{ex}}^{\text{tot}}$ is then to be performed from $r = 0$ to $r_{\max}$ for the same reason. Note also that an integration from $r_{\min} = 0$ to $r_{\max}$ changes the direction of integration (with respect to $\sigma$) which introduces an extra minus sign since we summed up all slices from bottom to top. This choice is convenient for concentric, monotonically decreasing distributions. However, our discussion is evidently true for arbitrary distributions if the boundaries are properly taken into account. See Fig. (4) for the geometrical meaning of all contributions.

To find the proper derivatives of Eqs. (44-46) one can apply the rules given in section (3.3).

### 5.1. Recovering BK

BK's result for homoeoidal symmetry is easily re-evaluated from Eq. (46) if we take $\epsilon$ and $\varphi$ as constants.

**Fig. 4.** The different contributions for the deflection of a density distribution as given by Eq.(46) at a location $z$ are shown. The *in*ternal part gives the contribution of all slices *in* which $z$ lies. The *ex*ternal part then gives all contributions to which $z$ lies external and the remaining part is due to the *thick* slice given by the cut-off at $r_{\max}$

For a position inside the distribution ($r(z) < r_{\max}$) Equation (44–45) yield (restricting to the right half plane for brevity)

$$\bar{I}_{\text{ex}}^{\text{tot}} = -\frac{1-\epsilon^2}{2\epsilon} \int_0^{r(z)} dr_E \, \sigma'(r_E) \left( \bar{z} - \sqrt{\bar{z}^2 - 4\epsilon r_E^2} \right) \quad . \tag{47}$$

Integration by parts gives

$$\bar{I}_{\text{ex}}^{\text{tot}} = -\frac{1-\epsilon^2}{2\epsilon} \left( \bar{z} - \sqrt{\bar{z}^2 - 4\epsilon r(z)^2} \right) \sigma(r(z))$$
$$+ 2(1-\epsilon^2) \int_0^{r(z)} \frac{\sigma(r_E) r_E dr_E}{\sqrt{\bar{z}^2 - 4\epsilon r_E^2}} \tag{48}$$

and

$$\bar{I}_{\text{in}}^{\text{tot}} = -(z - \epsilon \bar{z}) \left[ \sigma(r_{\max}) - \sigma(r(z)) \right] \quad . \tag{49}$$

Adding the contribution of $\bar{I}(\sigma(r_{\max}))$ we see that all terms cancel (using the same argument as in the derivation of Eq. (17)), except the remaining part which is just Eq. (11) as obtained by Bourassa and Kantowski. Similar arguments apply for a location outside the distribution if the comments after Eq. (46) are taken into account.

### 5.2. Confocal analytical models

For models with confocal symmetry the foci of each isodensity are at the same place by definition, i. e. $f$ and $\varphi$ are

With $r_{min} \geq f/2$ we obtain for Eq. (45) and Eq. (44) (omitting Bray's correction for brevity) two expressions which are related to a density distribution which is *constant* from zero to $r_{min}$ and then follows $\sigma(r_E)$. The varying remaining central part could be described by the general method explained above or for special (homoeoidal) cases by the BK or KK approach.

$$\bar{I}_{ex}^{tot} = -\frac{1}{2}\left(\bar{z} - \sqrt{\bar{z}^2 - f^2}\right)$$
$$\times \int_{r_{min}}^{r(z)} dr_E \sigma'(r_E)\left(\frac{4r_E^2}{f^2} - \frac{f^2}{4r_E^2}\right) \quad (50)$$

and

$$\bar{I}_{in}^{tot} = -\int_{r(z)}^{r_{max}} dr_E \sigma'(r_E)\left(z - \frac{f^2}{4r_E^2}\bar{z}\right) \quad . \quad (51)$$

These integrals lead obviously to closed solutions for many density distributions. For a *pseudo singular isothermal sphere*

$$\sigma_{psi}(r_E) = \frac{2\sigma_0}{2r_E - f} \quad ; \quad r_E \geq \frac{f}{2} \quad (52)$$

we find, e.g.,

$$\bar{I}_{ex}^{tot} = \frac{\sigma_0}{2}\left(\bar{z} - \sqrt{\bar{z}^2 - f^2}\right)$$
$$\times \left[\frac{(4r(z)r_{min} - f^2)(r(z) - r_{min})}{f^2 r(z) r_{min}}\right.$$
$$\left. + \frac{4}{f}\ln\left(\left(\frac{2r(z) - f}{2r_{min} - f}\right)^2 \frac{r_{min}}{r(z)}\right)\right] \quad (53)$$

and

$$\bar{I}_{in}^{tot} = \sigma_0 z \frac{4(r_{max} - r(z))}{(2r_{max} - f)(2r(z) - f)}$$
$$+ \sigma_0 \bar{z}\left[\frac{f - 4r(z)}{r(z)(2r(z) - f)} - \frac{f - 4r_{max}}{r_{max}(2r_{max} - f)}\right.$$
$$\left. - \frac{4}{f}\ln\left(\frac{2r(z) - f}{2r_{max} - f}\frac{r_{max}}{r(z)}\right)\right] \quad . \quad (54)$$

## 6. Conclusion

We showed how to derive the lensing properties of an elliptical slice of constant density from previous results of BK or Paper I. These slices are taken as basic elements to construct numerically or analytically models for the deflection due to surface density distributions of general elliptical symmetry. The results can be applied to observational data, assuming a constant mass-to-light ratio, taking the properties of the isodensities (density, ellipticity, twist angle) as input. For analytical purposes general one-dimensional integrals are derived for the deflection and applied (as an example for a closed solution) to the two-dimensional pseudo singular isothermal sphere with confocal symmetry.

Frankfurt and *Scientific Computers*, Aachen for hard- and software support and U. Borgeest, A. Dent, R. Kayser, P. King, S.Refsdal, A. Kassiola, I. Kovner, M. Birkinshaw and the referee R. D. Blandford for interesting discussions and helpful comments on the manuscript. This work was supported by the *Deutsche Forschungsgemeinschaft, DFG* under Schr 417/1-1.